# Valley-selective chiral phonon replicas of dark excitons and trions in monolayer WSe$_2$

Erfu Liu[1], Jeremiah van Baren[1], Takashi Taniguchi[2], Kenji Watanabe[2],

Yia-Chung Chang[3], Chun Hung Lui[1]*

[1] Department of Physics and Astronomy, University of California, Riverside, CA 92521, USA.

[2] National Institute for Materials Science, Tsukuba, Ibaraki 305-004, Japan

[3] Research Center for Applied Sciences, Academia Sinica, Taipei 11529, Taiwan

* Corresponding author. Email: joshua.lui@ucr.edu

**Abstract:**

We observe a set of three replica luminescent peaks at ~21.4 meV below the dark exciton, negative and positive dark trions (or exciton-polarons) in monolayer WSe$_2$. The replica redshift energy matches the energy of the zone-center E"-mode optical phonons. The phonon replicas exhibit parallel gate dependence and same g-factors as the dark excitonic states, but follow the valley selection rules of the bright excitonic states. While the dark states exhibit out-of-plane transition dipole and valley-independent linearly polarized emission in the in-plane directions, their phonon replicas exhibit in-plane transition dipole and valley-dependent circularly polarized emission in the out-of-plane directions. Our results and symmetry analysis show that the K-valley dark exciton decays into a left-handed chiral phonon and a right-handed photon, whereas the K'-valley dark exciton decays into a right-handed chiral phonon and a left-handed photon. Such valley selection rules of chiral phonon replicas can be utilized to identify the valleys of the dark excitonic states and explore their chiral interactions with phonons.

Monolayer transition metal dichalcogenides (TMDs), such as $MoS_2$ and $WSe_2$, are two-dimensional (2D) semiconductors with remarkable excitonic properties [1-3]. They host tightly bound excitons at two time-reversal valleys (K, K'), where spin-orbit coupling splits each band into two subbands with opposite spins [4-14]. If the electrons and holes come from bands with the same electron spin, they form bright excitons with a short lifetime (<10 ps) and relatively weak valley polarization due to the intervalley exchange interaction [Fig. 1(a)] [15]. But if the electrons and holes come from bands with opposite spins, they form dark excitons with a long lifetime (>100 ps) and relatively strong valley polarization due to the suppression of radiative recombination and intervalley exchange interaction [Fig. 1(b)] [16, 17]. These excitons can further couple to the Fermi sea to form trions [14, 18] or exciton-polarons [19-21] (hereafter we denote all of them as "trions" for simplicity). In monolayer $WS_2$ and $WSe_2$, the dark excitonic states can accumulate a large population because their energy level lies below the bright excitonic level [16, 17, 22-31]. These distinctive properties make the TMD dark excitonic states much better candidates than bright excitonic states for exciton transport, condensation, and valleytronic applications [32-34].

Detecting and manipulating the valley pseudospin of the dark excitonic states is, however, challenging because the usual valley selection rules of bright excitons are not applicable for dark excitons. The bright excitons exhibit in-plane (IP) transition dipole and couple selectively to light with right-handed (left-handed) circular polarization in the K (K') valley [Fig. 1(a)] [35-38]. We can therefore conveniently access their valley pseudospin by light helicity. But the dark excitonic states exhibit out-of-plane (OP) transition dipole and couple to vertically polarized light for both valleys [Fig. 1(b)] [17, 22-31]. The lack of valley selection rules poses a great challenge to study the dark-state valley dynamics. In search for new valley selection rules, the high-order processes, such as the decay of a dark exciton into a photon and a phonon, may exhibit valley selectivity. Z. Li *et al* recently reported a phonon replica of neutral dark exciton in monolayer $WSe_2$ [39], but their results are insufficient to establish rigorous valley selection rules for dark excitons and trions.

In this Rapid Communication, we experimentally establish the valley and chirality selection rules for both dark excitons and trions in their chiral-phonon replica emission in monolayer $WSe_2$. Our experiment reveals a set of three replica luminescent peaks at ~21.4 meV below the dark exciton, negative and positive dark trions. The redshift energy (~21.4 meV) matches the energy of the zone-center E"-mode optical phonons in monolayer $WSe_2$ [40-42]. The replica emission exhibits parallel gate dependence and same g-factors as the original dark states, but follows distinct optical selection rules. While the original dark states exhibit OP dipole and valley-independent linearly polarized emission in the IP directions, their phonon replicas exhibit IP dipole and valley-dependent circularly polarized emission in the OP directions, similar to the characteristics of the bright excitonic states [Fig. 1(c)]. Our results and symmetry analysis show that the K-valley dark exciton decays into a left-handed chiral phonon and a right-handed photon, whereas the K'-valley dark exciton decays into a right-handed chiral phonon and a left-handed photon. The replica PL intensity can be accounted for by first-principles calculations. Such valley-selective chiral phonon replicas can be utilized to identify the valley pseudospin of the dark excitonic states and explore the chiral exciton-phonon interactions.

We measure the photoluminescence (PL) from ultraclean monolayer WSe$_2$ devices encapsulated by boron nitride on Si/SiO$_2$ substrates under continuous 532-nm laser excitation at temperature T ~ 4 K [43]. Although the dark-state PL propagates in the IP directions, we can partially capture it with a wide-angle microscope objective (NA = 0.6) in the OP direction [16, 28, 43-47]. Figure 2a displays a gate-dependent PL map. The exceptional quality of our device allows us to observe a panoply of emission features, including the bright A exciton ($A^0$) and trions ($A_1^-$, $A_2^-$, $A^+$), dark exciton ($D^0$) and trions ($D^-$, $D^+$) [16, 17]. Notably, a PL peak emerges at ~21.4 meV below each of the $D^0$, $D^-$, $D^+$ peaks [Fig. 2(a-b)]. We denote them as $D_p^0$, $D_p^-$, $D_p^+$, respectively. We have fit the spectra with multiple Lorentzian functions and extracted their PL intensity and energy. Both their intensity and energy shift show similar gate dependence as the $D^0$, $D^-$, $D^+$ peaks [Fig. 2(c-d)]. We can visualize the parallel gate dependence in a second-derivative PL map ($d^2I/dE^2$), where the replica features are much sharpened [Fig. 2(e)].

The $D_p^0$, $D_p^-$, $D_p^+$ peaks exhibit almost the same g-factors as the $D^0$, $D^-$, $D^+$ peaks when we measure their Zeeman effect under out-of-plane magnetic field (B). The magnetic field can induce different band gaps in the two valleys [48-51]. The difference between the two valley gaps is the valley Zeeman splitting energy $\Delta E = g\mu_B B$, where g is the effective g-factor and $\mu_B = 57.88 \ \mu eV/T$ is the Bohr magneton. Figure 3(a-f) displays the B-dependent second-derivative PL maps (the raw data is in the Supplementary Materials [43]). From the linear Zeeman shift we can extract the g-factors. The $D_p^0$, $D_p^-$, $D_p^+$ peaks and the $D^0$, $D^-$, $D^+$ peaks have almost the same g-factors between -9.2 and -9.9 [Fig. 3(h)].

The parallel gate dependence and same g-factors strongly indicate that $D_p^0$, $D_p^-$, $D_p^+$ are the replicas of the $D^0$, $D^-$, $D^+$ peaks. We assign the $D_p^0$, $D_p^-$, $D_p^+$ peaks as the E"-mode phonon replicas because the redshift energy (~21.4 meV) matches the E" optical phonon energy in monolayer WSe$_2$ [40, 52-55]. Similar E" phonon replicas have been reported in quantum-dot excitons in monolayer WSe$_2$ [40]. In our experiment, the E" phonon replica is found only for the dark excitonic states, not for the bright excitonic states.

Although $D_p^0$, $D_p^-$, $D_p^+$ are the replicas of the dark states, they appear to follow the optical selection rules of the bright states. In the magneto-PL experiment for Figure 3, we excite the sample with linearly polarized laser and detect the PL with right- or left-handed helicity. Such measurements detect the dark states from both valleys because they emit linearly polarized light. But they detect the bright states only from one valley because the bright states emit right-handed (left-handed) light from the K (K') valley. Correspondingly, in our PL maps the $D^0$, $D^-$, $D^+$ peaks are each split into two branches under magnetic field, corresponding to the two valleys [Fig. (a-f)]. But the bright states only show Zeeman shift with no splitting, because we can only detect one valley. Remarkably, the three replica peaks exhibit the same behavior as the bright states – they also only show Zeeman shift with no splitting. In the right-handed PL detection, they shift in parallel with the lower branch of the dark states; we only observe the phonon replicas from the K valley. In the left-handed PL detection, they shift in parallel with the higher branch of the dark states; we only observe the phonon replicas from the K' valley. The helicity of the phonon replicas therefore tells us the valley pseudospin of the original dark states.

We have further obtained the transition dipole orientation of the phonon replicas. Y. Tang *at el* recently developed a special method to measure the dipole direction of the excitonic emission [17]. They deposit monolayer WSe$_2$ on a planar GaSe waveguide, which collects the light emission in the IP directions from both the IP and OP dipoles in monolayer WSe$_2$. By measuring the polarization of such emission, they can resolve the PL components from the IP and OP dipole [43]. We have extracted the exciton and replica PL intensity from their data and plot them as a function of polarization angle in Fig. 4. The dark and bright states exhibit perpendicular PL polarizations because they have OP and IP dipole, respectively. Notably, the phonon replicas have the same polarization as the bright states, indicating that they are associated with IP dipole.

The optical selection rules and IP dipole of the phonon replicas can be explained by the symmetry of the electron and phonon states by group theory (see Supplementary Materials for details [43]). The electronic states at the K/K' point possess the $C_{3h}$ symmetry point group, including the OP mirror symmetry ($\hat{\sigma}_h$) and IP three-fold rotation symmetry ($\hat{C}_3$). An eigenfunction $\psi$ transforms as $\hat{C}_3 \psi = e^{-i\frac{2\pi}{3}C_3} \psi$ and $\hat{\sigma}_h \psi = \sigma_h \psi$, where $C_3$ and $\sigma_h$ are the respective $\hat{C}_3$ and $\hat{\sigma}_h$ quantum numbers for $\psi$. Table 1 lists these quantum numbers for the spin-up and spin-down conduction bands ($c_\uparrow$, $\bar{c}_\downarrow$) and the spin-up valence band ($v_\uparrow$) at the K point. The hat on $\bar{c}_\downarrow$ denotes that it is not purely spin-down, but contains a small spin-up component from coupling to a higher spin-up band by spin-orbit coupling [49, 56]. Such spin mixing is necessary for the dark exciton to emit light. Under the symmetry operations, the $v - c$ interband transition matrices transform as [56, 57]:

$$\langle c|\hat{p}_\pm|v\rangle = \langle c|\hat{C}_3^{-1}\hat{C}_3 \hat{p}_\pm \hat{C}_3^{-1}\hat{C}_3|v\rangle = e^{i\frac{2\pi}{3}(C_3(c)-C_3(v)\mp 1)}\langle c|\hat{p}_\pm|v\rangle,$$

$$\langle c|\hat{p}_\pm|v\rangle = \langle c|\hat{\sigma}_h^{-1}\hat{\sigma}_h \hat{p}_\pm \hat{\sigma}_h^{-1}\hat{\sigma}_h|v\rangle = \sigma_h^*(c)\sigma_h(v)\langle c|\hat{p}_\pm|v\rangle.$$

$$\langle c|\hat{p}_z|v\rangle = \langle c|\hat{C}_3^{-1}\hat{C}_3 \hat{p}_z \hat{C}_3^{-1}\hat{C}_3|v\rangle = e^{i\frac{2\pi}{3}(C_3(c)-C_3(v))}\langle c|\hat{p}_z|v\rangle,$$

$$\langle c|\hat{p}_z|v\rangle = \langle c|\hat{\sigma}_h^{-1}\hat{\sigma}_h\hat{p}_z\hat{\sigma}_h^{-1}\hat{\sigma}_h|v\rangle = -\sigma_h^*(c)\sigma_h(v)\langle c|\hat{p}_z|v\rangle. \tag{1}$$

Here the momentum operators $\hat{p}_\pm = \hat{p}_x \pm i\hat{p}_y$ and $\hat{p}_z$ are associated with the IP chiral dipole and OP dipole, respectively. They transform as $\hat{C}_3\hat{p}_\pm\hat{C}_3^{-1} = e^{\mp i\frac{2\pi}{3}}\hat{p}_\pm$, $\hat{\sigma}_h\hat{p}_\pm\hat{\sigma}_h^{-1} = \hat{p}_\pm$, $\hat{C}_3\hat{p}_z\hat{C}_3^{-1} = \hat{p}_z$ and $\hat{\sigma}_h\hat{p}_z\hat{\sigma}_h^{-1} = -\hat{p}_z$. For a matrix element to be finite, the pre-factor after symmetry transformation must be one. From Table 1, we can verify that only $\langle c_\uparrow|\hat{p}_+|v_\uparrow\rangle$ and $\langle \bar{c}_\downarrow|\hat{p}_z|v_\uparrow\rangle$ can be finite, whereas other transition matrix elements are all zero. Therefore, the bright and dark excitons in the K valley are coupled exclusively to right-handed light and vertically polarized light, respectively [28, 35-38].

When the atoms move due to the lattice vibration, the original states are no longer eigenstates. In particular, the electron-phonon coupling will renormalize the $\bar{c}_\downarrow$ band into:

$$|\bar{\bar{c}}_\downarrow\rangle = |\bar{c}_\downarrow\rangle + \frac{\langle c_\uparrow,\Omega|\hat{H}_{ep}|\bar{c}_\downarrow\rangle}{E_{\bar{c}_\downarrow}-E_{c_\uparrow}-\hbar\Omega}|c_\uparrow\rangle. \tag{2}$$

Here $|\Omega\rangle$ denotes an E'' phonon with frequency $\Omega$. In the chiral mode, the W atoms stay stationary and the Se atoms rotate counter-clockwise or clockwise, giving rise to right-handed ($\Omega^+$) or left-handed ($\Omega^-$) phonons [Fig. 5(a)]. These phonons have odd mirror parity to mix the bright and dark states with opposite mirror parity. The chiral phonons also have three-fold rotation symmetry with quantum numbers $C_3(\Omega^\pm) = \pm 1$ (Table 1). Upon a $\hat{C}_3$ rotation, the matrix element transforms as:

$$\langle c_\uparrow,\Omega|\hat{H}_{ep}|\bar{c}_\downarrow\rangle = e^{i\frac{2\pi}{3}(1+C_3(\Omega))}\langle c_\uparrow,\Omega|\hat{H}_{ep}|\bar{c}_\downarrow\rangle. \tag{3}$$

The matrix element can only be finite for the left-handed phonon with $C_3(\Omega^-) = -1$. Therefore, the K-valley dark exciton only emits the left-handed chiral phonon.

The dark exciton can obtain oscillator strength from the bright exciton through the electron-phonon coupling $\hat{H}_{ep}$ and recombine through the $\bar{\bar{c}}_\downarrow - v_\uparrow$ transition by the electron-light interaction $\hat{H}_{el}$. The Fermi's golden rule gives the transition rate [39]:

$$P_{\bar{c}_\downarrow-v_\uparrow} \propto |\langle v_\uparrow,\omega,\Omega|\hat{H}_{el}|\bar{\bar{c}}_\downarrow\rangle|^2$$

$$\propto \left|\frac{\langle v_\uparrow,\omega,\Omega|\hat{H}_{el}|c_\uparrow,\Omega\rangle\langle c_\uparrow,\Omega|\hat{H}_{ep}|\bar{c}_\downarrow\rangle}{E_{\bar{c}_\downarrow}-E_{c_\uparrow}-\hbar\Omega}\right|^2. \tag{4}$$

Here $|\omega\rangle$ denotes a photon with frequency $\omega$. $\langle v_\uparrow,\omega,\Omega|\hat{H}_{el}|c_\uparrow,\Omega\rangle$ corresponds to the matrix element $\langle v_\uparrow|\hat{p}|c_\uparrow\rangle$ for the bright-exciton transition. Therefore, the chiral phonon replica follows the intensity and selection rules of bright exciton.

By combining the phonon and photon selection rules, we conclude that the dark exciton emits left-handed chiral phonon and right-handed photon in the K valley. By the time-reversal symmetry, it emits right-handed chiral phonon and left-handed photon in the K' valley [Fig. 5(b)]. These selection rules still hold even after we include the excitonic effect, because the exciton Hamiltonian has the same symmetry as the states at the K/K' point. Our experimental results are fully consistent with these selection rules.

While the excitonic effect does not modify the selection rules, it can substantially enhance the intensity of the phonon replicas. In particular, the finite k-space extent of the exciton envelop functions allows coupling to phonons with finite momentum. Only one chiral component of these phonons will contribute to the transition, so the chiral phonon selection rules still hold. We have calculated the replica intensity in a full excitonic picture with the density functional theory. The calculated intensity ratio between the replica and dark exciton is $I_{D_p^0}/I_{D^0} \approx 0.02$. This is close to the experimental ratio ($I_{D_p^0}/I_{D^0} \approx 0.05$) after we correct the different collection efficiency for the IP and OP emission in our setup (see Supplementary Materials [43]). Our experiment shows different replica ratios for dark trions ($I_{D_p^+}/I_{D^+} \approx 0.02$; $I_{D_p^-}/I_{D^-} \approx 0.07$). This indicates that the coupling to the Fermi sea can affect the phonon replica emission.

In summary, we have observed chiral phonon replicas of dark excitons and trions in monolayer WSe$_2$. The replicas exhibit rigorous chirality and valley optical selection rules, which allow us to access the dark-state valley pseudospin. The valley-selective replica emission can potentially be utilized to explore the valley dynamics of dark excitons and trions, such as to image their valley Hall effect. The replica process may also be used to generate phonons with selective chirality for the exploration of novel chiral phonon physics.

**ACKNOWLEDGMENTS**

We thank Y. Tang, K. F. Mak and J. Shan for sharing their dipole-resolved PL data with us. We thank D. Smirnov and Z. Lu for assistance in the magneto-optical experiment. A portion of this work was performed at the National High Magnetic Field Laboratory, which is supported by the National Science Foundation Cooperative Agreement No. DMR-1644779 and the State of Florida. Y.C.C. thanks C. T. Liang for assistance in the numerical calculation. Y.C.C. is supported by Ministry of Science and Technology (Taiwan) under grant no. MOST 107-2112-M-001-032. K.W. and T.T. acknowledge support from the Elemental Strategy Initiative conducted by the MEXT, Japan and the CREST (JPMJCR15F3), JST.

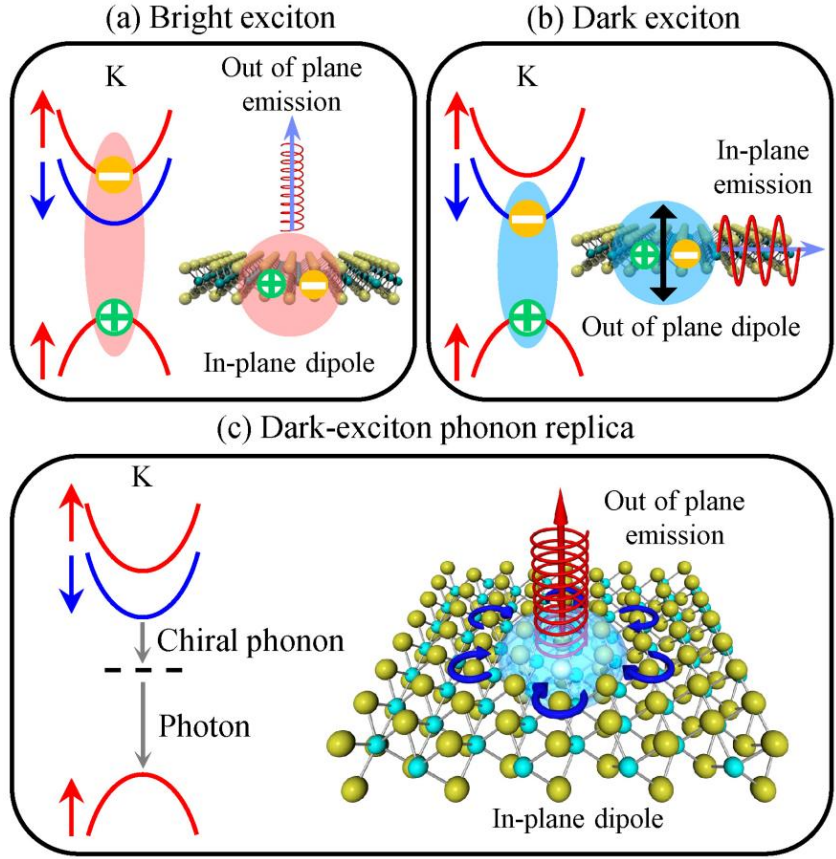

**FIG. 1**. (a-c) Band configurations, transition dipole, and optical emission of (a) bright exciton, (b) dark exciton, and (c) dark-exciton chiral phonon replica at the K valley in monolayer $WSe_2$. The arrows denote the electron spin. A dark exciton can decay into a chiral phonon and a photon with opposite chirality.

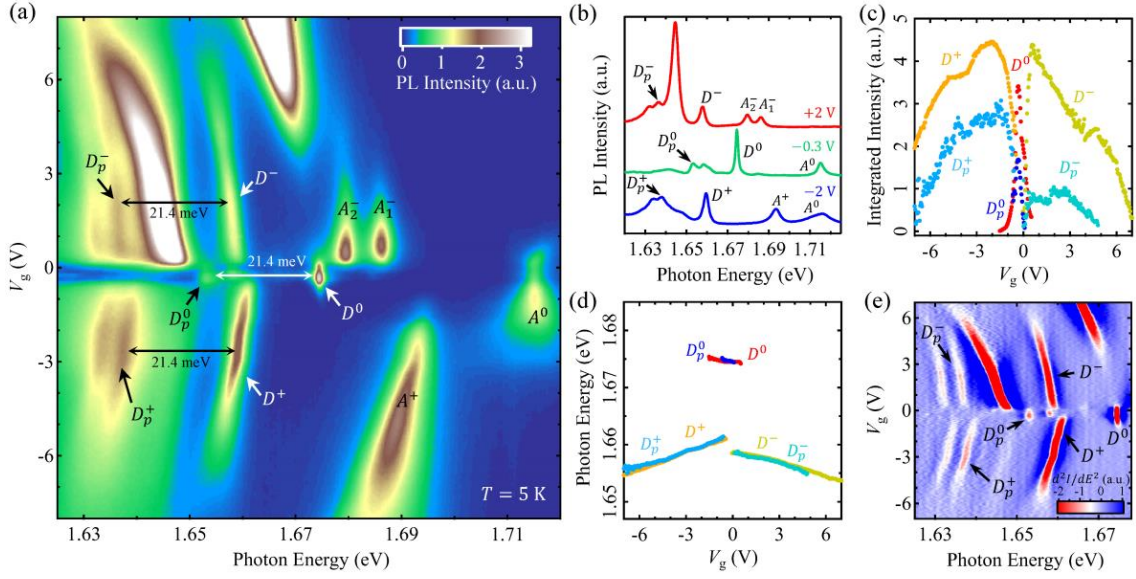

**FIG. 2**. (a) Gate-dependent photoluminescence (PL) map of a monolayer WSe$_2$ device encapsulated by boron nitride. We denote the bright excitonic states ($A^0$, $A^+$, $A_1^-$, $A_2^-$), dark excitonic states ($D^0$, $D^-$, $D^+$), and dark states phonon replicas ($D_p^0$, $D_p^-$, $D_p^+$). (b) The cross-cut PL spectra at the charge neutrality point (gate voltage $V_g = -0.3$ V), electron side ($V_g = 2$ V), and hole side ($V_g = -2$ V). (c) PL intensity and (d) PL photon energy of the dark excitonic states and replicas as a function of gate voltage. The replica energy is upshifted for 21.4 meV for comparison. (e) The second energy derivative ($d^2I/dE^2$) of the PL map in panel (a).

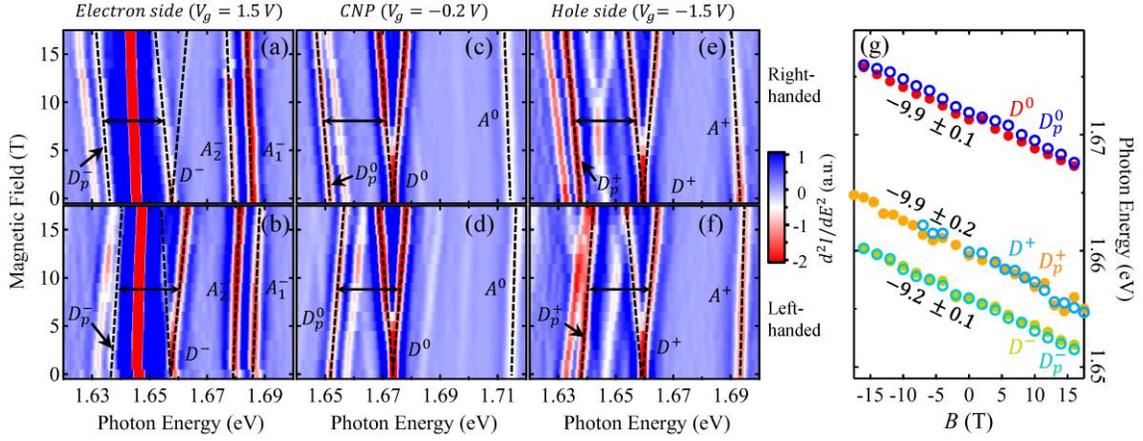

**FIG. 3**. (a-f) Magnetic-field dependent second-derivative PL map ($d^2I/dE^2$) of monolayer WSe$_2$ on the electron side (a-b; gate voltage $V_g$ = 1.5 V), near the charge neutrality point (CNP) (c-d; $V_g$ = –0.2 V) and on the hole side (e-f; $V_g$ = –1.5 V). We excite the sample with linearly polarized 532-nm laser and collect the PL with right-handed (top row) or left-handed helicity (bottom row). (g) The Zeeman energy shift of one branch of the dark states and their replicas. The replica energy is upshifted for 21.4 meV for comparison. The Zeeman-splitting g-factors from linear fits are denoted.

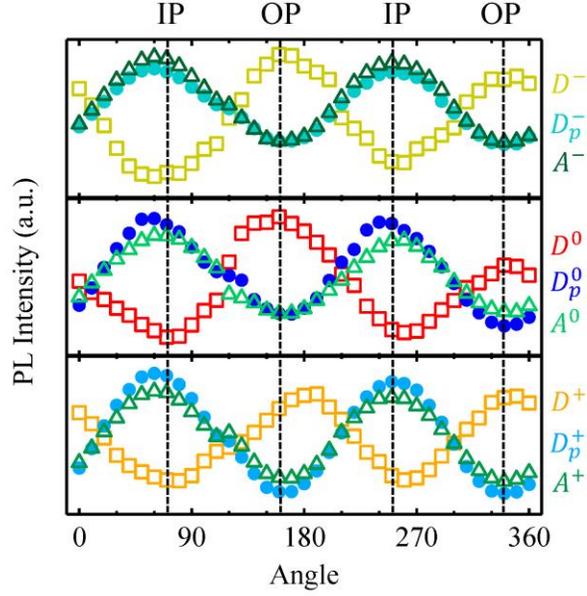

**FIG. 4**. The PL intensity of the bright excitonic states (triangle), dark excitonic states (square), and dark-state phonon replicas (dots) as a function of polarization angle in the in-plane collection geometry. The angles corresponding to in-plane (IP) and out-of-plane (OP) dipole are denoted.

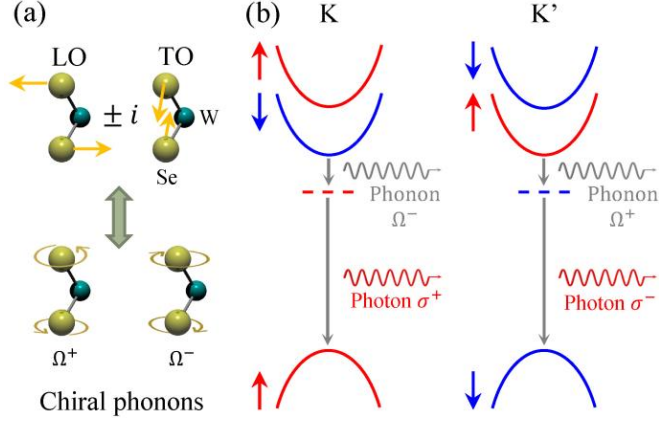

**FIG. 5.** (a) The configurations of the doubly-degenerate zone-center $E''$ phonons in monolayer WSe$_2$. The W atoms are stationary and the Se atoms move laterally. The vibration can be decomposed into the LO and TO modes with linear Se atomic motion or left-handed and right-handed chiral modes with rotational Se atomic motion. (b) The phonon-assisted radiative recombination of dark exciton. The dark exciton can decay into a pair of phonon and photon with opposite chirality.

|  | $c_\uparrow$ | $\bar{c}_\downarrow$ | $v_\uparrow$ | $\Omega^+$ | $\Omega^-$ |
|---|---|---|---|---|---|
| $\sigma_h$ | $i$ | $-i$ | $i$ | $-1$ | $-1$ |
| $C_3$ | $+\frac{1}{2}$ | $-\frac{1}{2}$ | $-\frac{1}{2}$ | $+1$ | $-1$ |

**Table 1.** Symmetry quantum numbers for the electronic bands ($c_\uparrow$, $\bar{c}_\downarrow$, $v_\uparrow$) at the K point and the E" chiral phonons ($\Omega^+$, $\Omega^-$) at the zone center for monolayer WSe$_2$.

# Supplementary Materials of
# Chiral-phonon replicas of dark excitonic states in monolayer WSe$_2$

Erfu Liu[1], Jeremiah van Baren[1], Zhengguang Lu[2,3], Takashi Taniguchi[4], Kenji Watanabe[4], Dmitry Smirnov[2], Yia-Chung Chang[5], Chun Hung Lui[1]*

[1] Department of Physics and Astronomy, University of California, Riverside, CA 92521, USA.

[2] National High Magnetic Field Laboratory, Tallahassee, FL 32310, USA

[3] Department of Physics, Florida State University, Tallahassee, FL 32310, USA

[4] National Institute for Materials Science, Tsukuba, Ibaraki 305-004, Japan

[5] Research Center for Applied Sciences, Academia Sinica, Taipei 11529, Taiwan

* Corresponding author. Email: joshua.lui@ucr.edu

**1. Supplementary information for experiments at zero magnetic field**
    1.1. Device fabrication
    1.2. Experimental methods
    1.3. Power dependence of photoluminescence
    1.4. Temperature dependence of photoluminescence
    1.5. Dipole-resolved photoluminescence experiment

**2. Supplementary information for experiments under magnetic field**
    2.1. Magneto-optical experimental methods
    2.2. Zeeman effect and g-factors

**3. Group theory of bright exciton, dark exciton and chiral-phonon replica**
    3.1. Electronic bands and selection rules without spin-orbit coupling
    3.2. Electronic bands and selection rules with spin-preserving spin-orbit coupling
    3.3. Electronic bands and selection rules with spin-flipping spin-orbit coupling
    3.4. Electronic bands and selection rules with electron-phonon coupling
    3.5. Possible photon-phonon entanglement from the decay of dark exciton

**4. First-principles calculations of the emission intensity**
    4.1 Intensity ratio between bright exciton and dark exciton
    4.2 Intensity ratio between dark exciton and phonon replica

# 1. Supplementary information for experiments at zero magnetic field

## 1.1. Device fabrication

Monolayer WSe$_2$ devices encapsulated by hexagonal boron nitride (BN) are fabricated by mechanical co-lamination of two-dimensional (2D) crystals. We use WSe$_2$ bulk crystals from HQ Graphene Inc. We first exfoliate monolayers WSe$_2$, multi-layer graphene and thin BN flakes from their bulk crystals onto the Si/SiO$_2$ substrates. Afterward, a polycarbonate-based dry-transfer technique is applied to stack the different 2D crystals together. We use a stamp to first pick up a BN flake, and sequentially pick up multi-layer graphene (as the electrodes), a WSe$_2$ monolayer, a BN thin layer (as the bottom gate dielectric), and a graphene multi-layer (as the back gate). This method ensures that the WSe$_2$ layer doesn't contact the polymer during the whole fabrication process, so as to reduce the contaminants and bubbles at the interface. Afterward, standard electron beam lithography is applied to pattern and deposit the gold contacts (100 nm thickness). Finally, the devices are annealed at 300 °C for 3 hours in an argon environment.

Fig. S1(a-b) display the schematic of our devices and the optical image of a representative device. For Device 1 used in Fig. 2–3 of the main paper, the bottom BN gate dielectric layer is about 25 nm in thickness, from which we can deduce the capacity and gating charge density of the back gate.

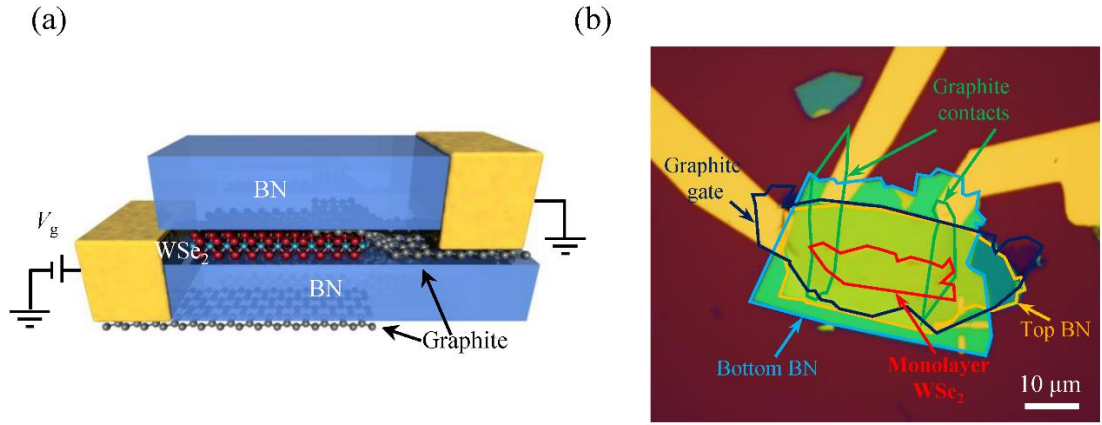

**Fig. S1. (**a) The schematic of BN-encapsulated monolayer WSe$_2$ device. (b) The optical image of a representative device.

## 1.2. Experimental methods

The experiments under no magnetic field (Fig. 2, S2, S3) were conducted in our lab at UC Riverside. The samples were mounted inside a helium-cooled cryostat (Montana). We excited the samples with a 532-nm continuous laser (Laser Quantum; Torus 532). The laser was focused through a microscope objective (40X, NA = 0.6). The spot diameter on the sample was ~1 μm. The photoluminescence (PL) was collected through the same objective in a backscattering geometry. The PL spectra were analyzed by a high-resolution spectrometer with a charge-coupled device (CCD) camera (Princeton Instruments).

## 1.3. Power dependence of photoluminescence

We have measured the PL from bright excitons ($A^0$), dark excitons ($D^0$) and phonon replica of dark excitons ($D_p^0$) on Device 1 at different excitation laser power (Fig. S2). They all exhibit approximately linear power dependence. The power-law fits $I = P^\gamma$ give the exponents $\gamma$ = 1.15, 1.06 and 0.99 for the $A^0$, $D^0$ and $D_p^0$ peaks, respectively. The linear power dependence shows that the $D_p^0$ peak is not associated with the biexciton state with quadratic power dependence.

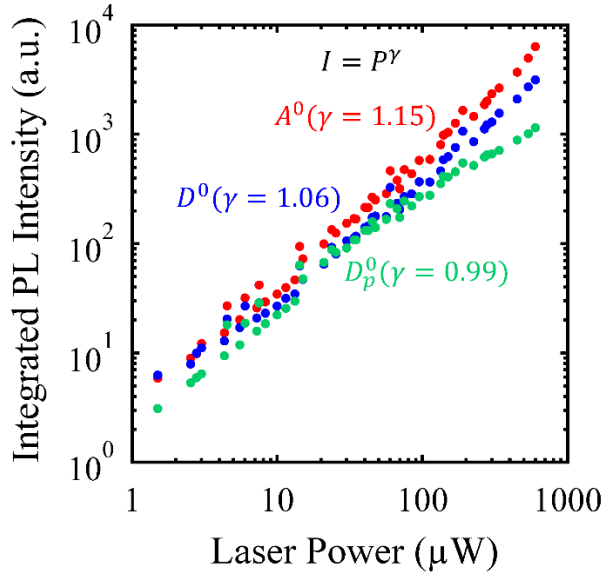

**Fig. S2.** PL intensity as a function of incident laser power for the bright exciton ($A^0$, red), dark exciton ($D^0$, blue), phonon replica of dark exciton ($D_p^0$, green). The lines are fits by the power-law $I = P^\gamma$.

## 1.4. Temperature dependence of photoluminescence

We have measured the PL spectra of dark excitonic states and their phonon replicas at different temperatures from 5 to 50 K (Fig. S3). As the temperature decreases, the bright exciton peak becomes weaker, but the dark exciton peak grows stronger, as observed also in prior experiments [1, 2]. The result can be understood from the different energy levels of bright and dark excitons. In monolayer WSe$_2$, the bright excitons occupy an elevated energy level. Their population (and hence the PL intensity) decreases at low temperature, when the thermal energy is not sufficient to sustain the high-lying bright excitons. In contrast, the dark excitons occupy the lowest energy level. Their population accumulates and increases at low temperature, leading to stronger PL at low temperature. The similar behavior is observed for dark exciton-polarons when we change the gate voltage. The intensity of the $D_p^0$, $D_p^+$, $D_p^+$ peaks also increases at low temperature, with similar tendency as the dark excitonic states. The similar temperature dependence supports our assignment that the $D_p^0$, $D_p^+$, $D_p^+$ peaks are the replicas of the dark excitonic states.

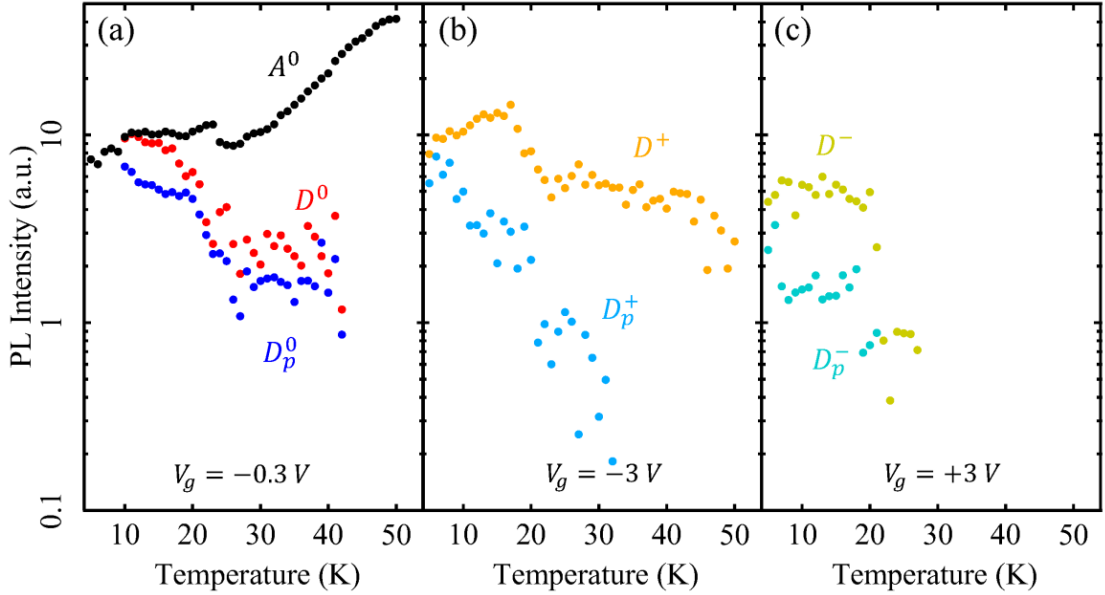

**Fig. S3.** (a) The integrated PL intensity of bright exciton ($A^0$), dark exciton ($D^0$) and dark-exciton phonon replica ($D_p^0$) as a function of temperature. (b-c) The integrated PL intensity of positive and negative dark exciton-polaron ($D^+$, $D^-$) and their replicas ($D_p^+$, $D_p^-$) as a function of temperature. The gate voltages ($V_g$) are denoted in the figure.

### 1.5. Dipole-resolved photoluminescence experiment

In Fig. 4 of the main paper, we have used a waveguide-based method to measure the dipole orientation of the PL emission. The method was developed by Yanhao Tang *et al* and the details can be found in Ref. [3]. The experiment uses a special device geometry, in which the BN-encapsulated monolayer WSe$_2$ device is deposited upon a GaSe layer with a thickness of several hundred nanometers. The GaSe layer, acting as a waveguide, collects the light emission from monolayer WSe$_2$ along the in-plane direction. After the light propagates along the GaSe layer, some of the light will be deflected upward at the edge of the GaSe layer. By using a half-wave plate and a linear polarizer, they can selectively detect the component of the PL along any linear polarization direction. The in-plane PL can come from both the in-plane (IP) dipole and out-of-plane (OP) dipole. The polarizations of emission from the IP and OP dipoles are expected to remain largely perpendicular to each other after propagating shortly through the GaSe waveguide. Therefore, the polarization-resolved PL map allows us to identify the dipole direction of different peaks in the PL spectra.

## 2. Supplementary information for experiments under magnetic field

### 2.1. Magneto-optical experimental methods

The experiments under magnetic field were performed in the National High Magnetic Field Laboratory in Tallahassee, Florida. We used a 17.5-T DC magnet. Fig. S4 displays the schematic of the experimental setup. We used a 532-nm continuous laser as the excitation light source. The laser beam was focused by a lens (NA = 0.67) onto the sample. The sample was mounted on a three-dimensional piezoelectric translation stage. The PL was collected through a 50/50 beam splitter into a multimode optical fiber, and subsequently measured by a spectrometer with a CCD camera (Princeton Instruments, IsoPlane 320). A quarter wave plate and a polarization beam splitter were used to select the right-handed or left-handed circularly polarized component of the PL signal.

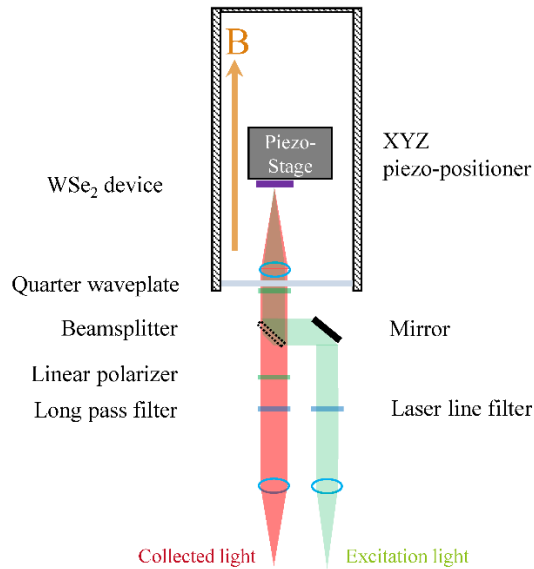

**Fig. S4.** Magneto-optical experimental setup.

### 2.2. Zeeman effect and g-factors

The application of out-of-plane magnetic field can break the time-reversal symmetry and lift the degeneracy of the K and K' valleys in monolayer WSe$_2$. The magnetic field can increase the energy gap of one valley and diminish the energy gap of the other valley. The difference between the two valley gaps is defined as the valley Zeeman splitting energy $\Delta E = g\mu_B B$, where $\mu_B$ is the Bohr magneton, $B$ is the magnetic field, and $g$ is the effective g-factor. The valley Zeeman splitting energy can be measured from the PL of two valleys under magnetic field or from the PL of one valley under opposite magnetic field [4-7]. We have measured the helicity-resolved PL spectra under magnetic field. Fig. 3 of the main paper displays the second energy derivative of the B-dependent PL maps. We take the second derivative because it can sharpen the weak features and help us identify their PL energies. Here in Fig. S5, we show both the original and second-derivative PL maps for comparison. We denote the g-factors of various observed peaks on the top of Fig. S5. Fig. S6 displays the cross-cut PL spectra from Fig. 5(b).

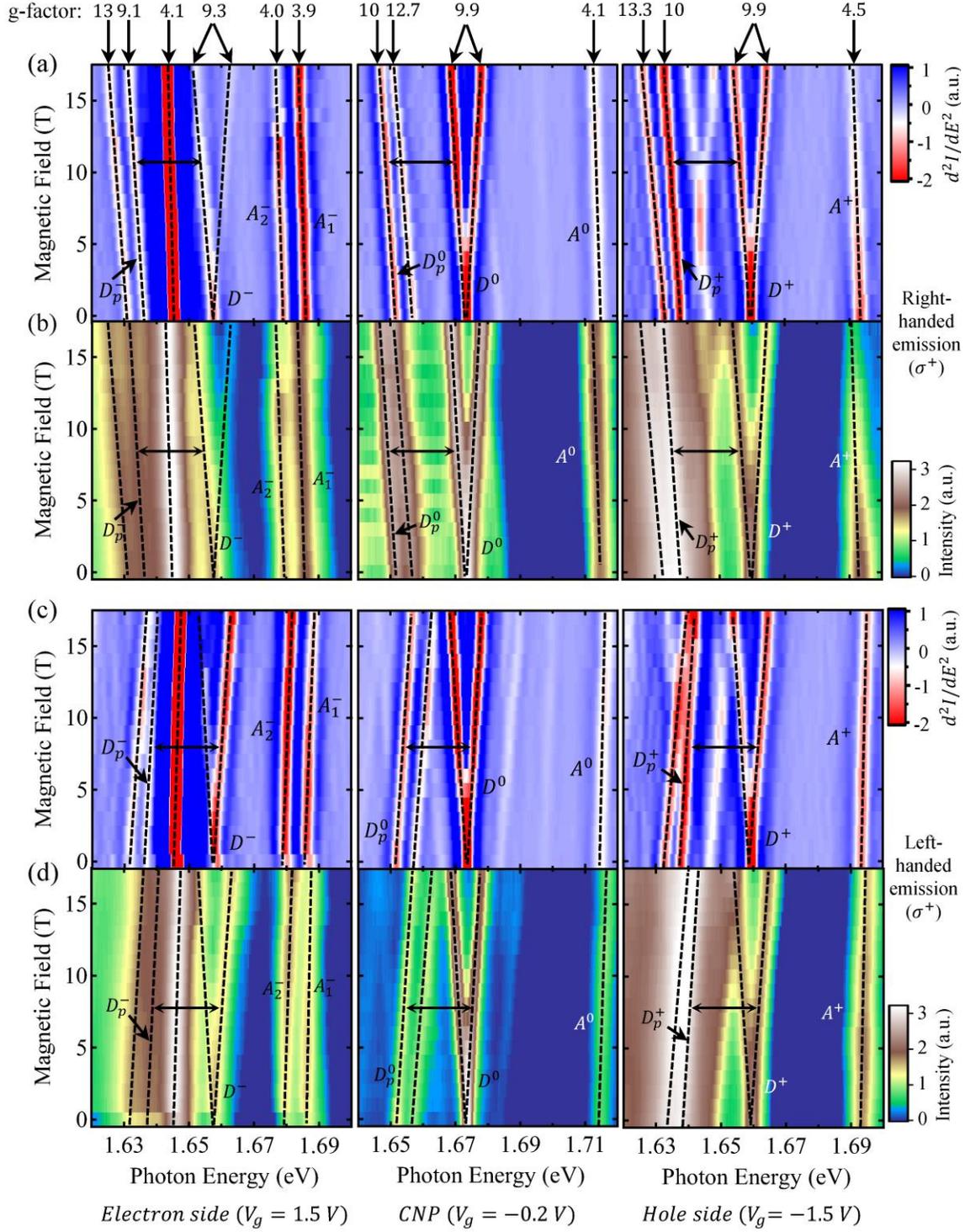

**Fig. S5.** (a-d) Magnetic-field dependent PL map of monolayer WSe$_2$ on the electron side (first column; $V_g = 1.5$ V), near the charge neutrality point (second column; $V_g = -0.2$ V) and on the hole side (third column; $V_g = -1.5$ V). We excite the sample with linearly polarized 532-nm laser and collect the PL with right-handed helicity ($\sigma^+$; a, b) and left-handed helicity ($\sigma^-$; c, d). The top row denotes the g-factors of the observed peaks. Panels (a, c) are the second energy derivative of panels (b, d), respectively.

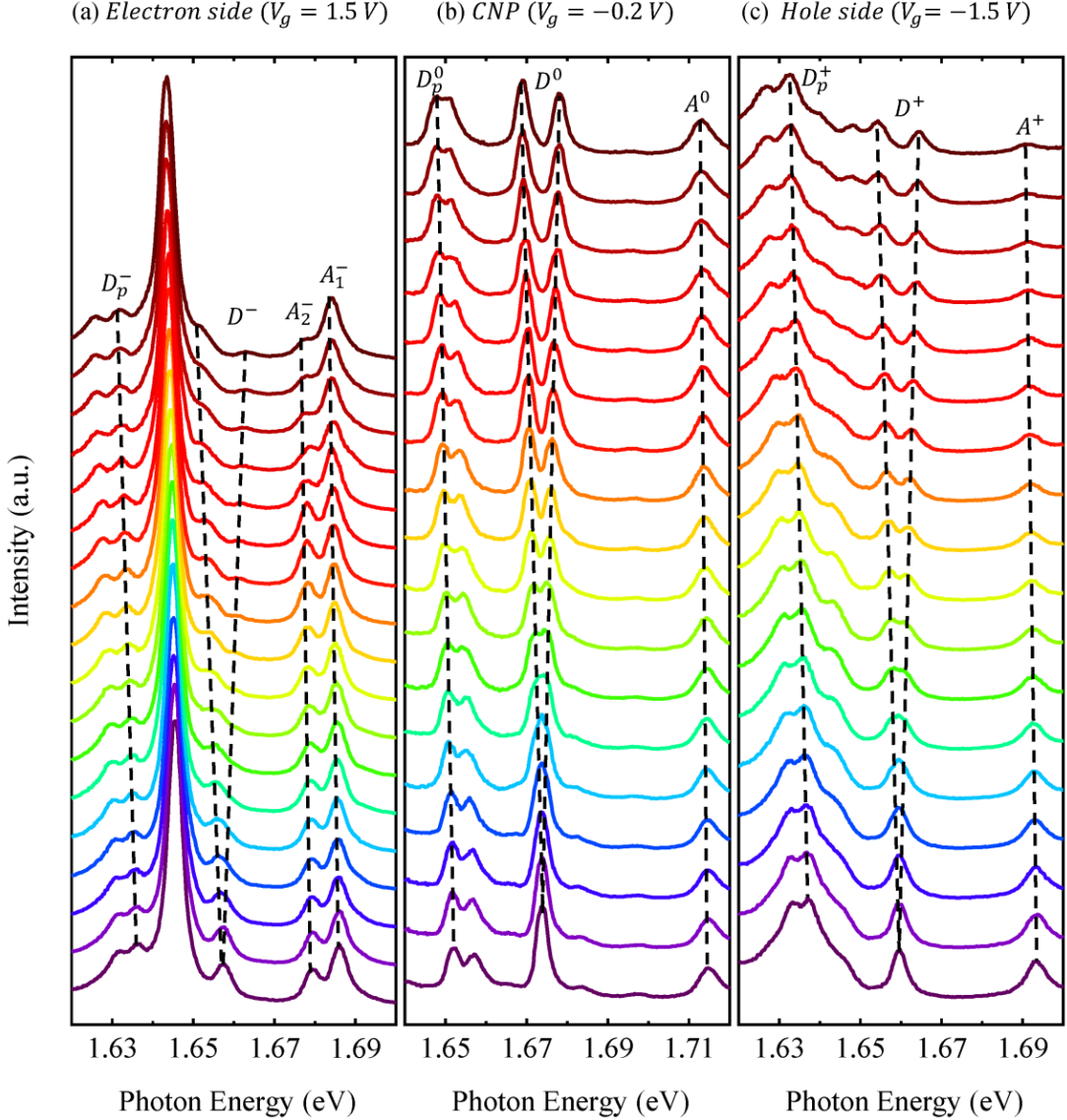

**Fig. S6.** (a-c) Cross-cut PL spectra of the right-handed PL maps in Fig. S5(b). The magnetic field increases from 0 to 17 T from bottom to top with 1 T even increment. The bottom line is the spectra at zero magnetic field. The dashed lines highlight the valley Zeeman shifts.

Besides the excitonic states, and replica peaks, we also observe a few other weak peaks with g-factors of about –13. The different g-factors show that they are not the replica peaks of bright or dark excitonic states. They may come from the momentum-indirect excitonic states in monolayer $WSe_2$. Further research is merited to clarify the origins of these peaks.

## 3. Group theory of bright exciton, dark exciton and chiral-phonon replica

We are able to explain the emission polarization and dipole orientation of bright exciton, dark exciton and chiral-phonon replicas from the symmetry of electronic states and phonon states. In this section, we will derive their transition selection rules by group theory. We will consider only the K point here. The selection rules at the K' point can be obtained straightforwardly by using time-reversal symmetry. Also, we can neglect the excitonic effect here. Although the exciton consists of states near the K point, the exciton has exactly the same selection rules of the transitions at the K point, because the exciton Hamiltonian at the K valley has the same symmetry of the states at the K point. Therefore, it is sufficient to consider only the free electronic states at the K point to explain the selection rules of excitons.

For the sake of clarity, we will present the group theory in four progressing stages, as illustrated in Fig. S7. First, we will consider the electronic bands with no spin-orbit coupling (SOC). This simple picture is sufficient to derive the optical selection rules of bright excitons.  Second, we will consider the influence of spin-preserving SOC, which gives rise to the spin-forbidden dark excitons.  Third, we will consider the influence of spin-flipping SOC, which enables the dark excitons to emit z-polarized light.  Finally, we will consider the influence of electron-phonon coupling, which mixes the bright and dark states and enables the chiral-phonon-assisted optical recombination of dark excitons.

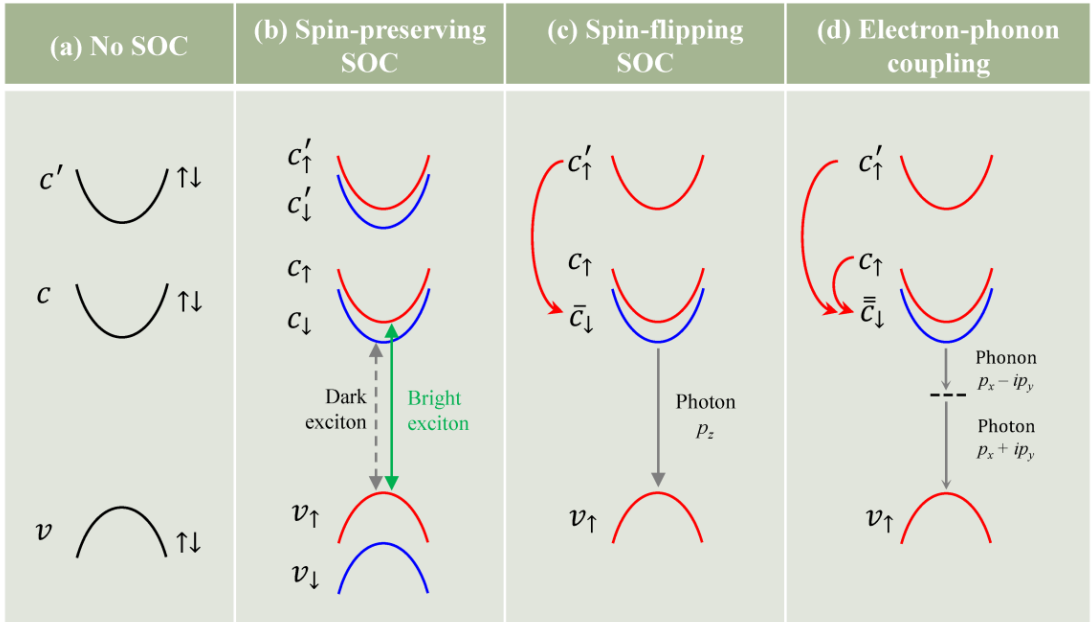

**Fig. S7.** Band configurations for four progressive interaction pictures in monolayer WSe$_2$. (a) Electronic bands with no spin-orbit coupling (SOC). (b) Electronic bands with the spin-preserving SOC. Each band is split to the two bands with spin-up and spin-down configurations. (c) Electronic bands including the spin-flipping SOC terms, which mix the $c'_\uparrow$ and $c_\downarrow$ band into the $\bar{c}_\downarrow$ band (red arrow). The subscript "$\downarrow$" in $\bar{c}_\downarrow$ only means the spin state is mostly (not entirely) spin-down. For simplicity, we only show four bands relevant to our analysis here, whereas other bands are removed. (d) Electronic bands with electron-phonon coupling, which mix the $c_\uparrow$ and $\bar{c}_\downarrow$ bands into the $\bar{\bar{c}}_\downarrow$ band (red arrow). The bright/dark excitons, phonon replica, and their transition dipole polarizations are denoted.

### 3.1. Electronic bands and selection rules without spin-orbit coupling

We first consider the electronic bands at the K valley without including the SOC.  For our purpose of understanding the dark-exciton phonon replica, it is sufficient to consider only three bands – the highest valence band ($v$), the lowest conduction band ($c$) and the next-lowest conduction band ($c'$, ~1 eV higher than $c$) [Fig. S7(a)].  In the absence of SOC, each of these bands has the two-fold spin degeneracy.  According to prior DFT calculations, they are all dominated by the d-orbits at the W atoms [8-10].  The p-orbits at the Se atoms have a small contribution, which can be ignored in our analysis here.  The prismatic crystal field at the W atom splits the d-orbits into three levels – the $d_0$, $d_{\pm 1}$, $d_{\pm 2}$ orbits, which are defined as:

$$d_0 = d_{z^2},$$
$$d_{\pm 1} = \tfrac{1}{\sqrt{2}}(d_{xz} \pm i d_{yz}),$$
$$d_{\pm 2} = \tfrac{1}{\sqrt{2}}(d_{x^2-y^2} \pm i d_{xy}). \tag{S1}$$

The subscripts $0$, $\pm 1$, $\pm 2$ denote the magnetic angular momentum quantum number ($m$). The $c'$, $c$ and $v$ bands at the K valley are dominated by the $d_{-1}$, $d_0$ and $d_{+2}$ orbits respectively (Table S1).  The corresponding bands at the K' valley are dominated by the d-orbits with opposite $m$.

| Bands | W d-orbits | $\sigma_h$ | $C_3(W)$ | $C_3(Se)$ | $C_3(h)$ |
|---|---|---|---|---|---|
| $c'$ | $d_{-1}$ | $-1$ | $-1$ | $+1$ | $0$ |
| $c$ | $d_0$ | $+1$ | $0$ | $-1$ | $+1$ |
| $v$ | $d_{+2}$ | $+1$ | $-1$ | $+1$ | $0$ |

**Table S1.** Dominant orbital components and symmetry quantum numbers for the highest valence band ($v$), the lowest conduction band ($c$) and the next-lowest conduction band ($c'$) at the K point in monolayer WSe$_2$ without considering the spin and spin-orbit coupling.  $\sigma_h$ is the out-of-plane mirror parity. $C_3(W)$, $C_3(Se)$, and $C_3(h)$ are the quantum numbers for the 120-degree in-plane rotation around the W atom, Se atom, and the hexagon center ($h$), respectively.  For the time-reversal K' point, $\sigma_h$ remains the same but all $C_3$ numbers become opposite in sign.

The electronic states at the K point possess the $C_{3h}$ symmetry point group, including the out-of-plane mirror symmetry ($\hat{\sigma}_h$) and in-plane three-fold rotation symmetry ($\hat{C}_3$). The $\hat{C}_3$ operation means rotating the wavefunction (not coordinate) counter-clockwise by 120 degree. An orbital eigenfunction $\psi$ (with no spin) transforms as:

$$\hat{C}_3 \psi = e^{-i\frac{2\pi}{3}C_3(\psi)} \psi,$$

$$\hat{\sigma}_h \psi = \sigma_h(\psi) \psi. \tag{S2}$$

Here $C_3(\psi) = 0, \pm 1$ and $\sigma_h(\psi) = \pm 1$ are the respective $\hat{C}_3$ and $\hat{\sigma}_h$ quantum numbers for $\psi$. The symmetry quantum numbers for the $c'$, $c$ and $v$ bands at the K point are summarized in Table S1, which includes the $\hat{C}_3$ rotation around the W atom, Se atom and the hexagon center (h).

From Table S1 we can derive the selection rules for the optical transitions at the K point. The $v-c$ interband transitions are characterized by the optical matrix element $\langle v|\hat{\boldsymbol{p}}|c\rangle$, where $\hat{\boldsymbol{p}} = \hat{p}_+ \boldsymbol{e}_+^* + \hat{p}_- \boldsymbol{e}_-^* + \hat{p}_z \boldsymbol{z}$ is the momentum operator and $\boldsymbol{e}_\pm = (\boldsymbol{x} \pm i\boldsymbol{y})/\sqrt{2}$ and $\boldsymbol{z}$ are the coordinate unit vectors. $\hat{p}_\pm = \hat{p}_x \pm i\hat{p}_y$ are associated with the right-handed ($\sigma^+$) and left-handed ($\sigma^-$) circularly polarized light, respectively. $\hat{p}_z$ is associated with the $z$-polarized light. These momentum operators transform as:

$$\hat{C}_3 \hat{p}_\pm \hat{C}_3^{-1} = e^{\mp i\frac{2\pi}{3}} \hat{p}_\pm,$$

$$\hat{\sigma}_h \hat{p}_\pm \hat{\sigma}_h^{-1} = \hat{p}_\pm,$$

$$\hat{C}_3 \hat{p}_z \hat{C}_3^{-1} = \hat{p}_z,$$

$$\hat{\sigma}_h \hat{p}_z \hat{\sigma}_h^{-1} = -\hat{p}_z. \tag{S3}$$

The matrix elements then transform as:

$$\langle c|\hat{p}_\pm|v\rangle = \langle c|\hat{C}_3^{-1}\hat{C}_3\hat{p}_\pm\hat{C}_3^{-1}\hat{C}_3|v\rangle = e^{i\frac{2\pi}{3}(C_3(c)-C_3(v)\mp 1)}\langle c|\hat{p}_\pm|v\rangle,$$

$$\langle c|\hat{p}_\pm|v\rangle = \langle c|\hat{\sigma}_h^{-1}\hat{\sigma}_h\hat{p}_\pm\hat{\sigma}_h^{-1}\hat{\sigma}_h|v\rangle = \sigma_h^*(c)\sigma_h(v)\langle c|\hat{p}_\pm|v\rangle,$$

$$\langle c|\hat{p}_z|v\rangle = \langle c|\hat{C}_3^{-1}\hat{C}_3\hat{p}_z\hat{C}_3^{-1}\hat{C}_3|v\rangle = e^{i\frac{2\pi}{3}(C_3(c)-C_3(v))}\langle c|\hat{p}_z|v\rangle,$$

$$\langle c|\hat{p}_z|v\rangle = \langle c|\hat{\sigma}_h^{-1}\hat{\sigma}_h\hat{p}_\pm\hat{\sigma}_h^{-1}\hat{\sigma}_h|v\rangle = -\sigma_h^*(c)\sigma_h(v)\langle c|\hat{p}_z|v\rangle. \tag{S4}$$

For a matrix element to be finite, the pre-factor after the symmetry transformation has to be one. From Table S1, $C_3(c) - C_3(v) = 3N + 1$ and $\sigma_h^*(c)\sigma_h(v) = 1$, where $N$ is an integer. So, only $\langle c|\hat{p}_+|v\rangle$ can be finite, whereas $\langle c|\hat{p}_-|v\rangle$ and $\langle c|\hat{p}_z|v\rangle$ are both zero. Therefore, the $v-c$ transition at the K point is coupled only to the right-handed light (By time-reversal symmetry, the corresponding transition at the K' point is coupled only to the left-handed light). This is the famous chiral valley selection rules for the bright excitons in monolayer TMDs [11-15]. We note that the initial and final states must have the same spin (the condition of bright exciton), because photon cannot flip the electron spin.

## 3.2. Electronic bands and selection rules with spin-preserving spin-orbit coupling

We next consider the effect of the electron spin and spin-orbit coupling. Spin has very different symmetry characters from the orbits. The spin-up (spin-down) state has a mirror parity $\sigma_h = i(-i)$ and a rotation quantum number $C_3 = +\frac{1}{2}(-\frac{1}{2})$ (Table S2). It takes four times of mirror reflection or a two-round rotation to restore the spin state to itself.

| Spin | $\sigma_h$ | $C_3$ |
|---|---|---|
| $\|\uparrow\rangle$ | $i$ | $+\frac{1}{2}$ |
| $\|\downarrow\rangle$ | $-i$ | $-\frac{1}{2}$ |

Table S2. The symmetry quantum numbers for electron spin.

WSe$_2$ possesses strong spin-orbit coupling. The SOC Hamiltonian can be written as $\hat{H}_{SOC} = \alpha(\hat{L}_z\hat{S}_z + \hat{L}_+\hat{S}_- + \hat{L}_-\hat{S}_+)$, where $\alpha$ is a material parameter to characterize the SOC strength. The first term preserves the spin; the second and third terms flip the spin. We consider only the first term in this section and will consider the second and third terms in the next section. The spin-preserving term $\alpha\hat{L}_z\hat{S}_z$ splits each band into two subbands with spin-up and spin-down configuration [Fig. S7(b)]. For the $v$ band with the dominant $d_{+2}$ orbit, the SOC splitting energy reaches ~450 meV in monolayer WSe$_2$. For the $c$ band, the dominant $d_0$ orbit induces no SOC splitting; only the minority p-orbits at the Se atoms induce a small splitting of ~30 meV [9, 16, 17].

| Bands | W d-orbitals | $\sigma_h$ | $C_3(W)$ | $C_3(Se)$ | $C_3(h)$ |
|---|---|---|---|---|---|
| $c'_\uparrow$ | $d_{-1}$ | $-i$ | $-\frac{1}{2}$ | $+\frac{3}{2}$ | $+\frac{1}{2}$ |
| $c'_\downarrow$ |  | $i$ | $-\frac{3}{2}$ | $+\frac{1}{2}$ | $-\frac{1}{2}$ |
| $c_\uparrow$ | $d_0$ | $i$ | $+\frac{1}{2}$ | $-\frac{1}{2}$ | $+\frac{3}{2}$ |
| $c_\downarrow(\bar{c}_\downarrow)$ |  | $-i$ | $-\frac{1}{2}$ | $-\frac{3}{2}$ | $+\frac{1}{2}$ |
| $v_\uparrow$ | $d_{+2}$ | $i$ | $-\frac{1}{2}$ | $+\frac{3}{2}$ | $+\frac{1}{2}$ |
| $v_\downarrow$ |  | $-i$ | $-\frac{3}{2}$ | $+\frac{1}{2}$ | $-\frac{1}{2}$ |

Table S3. Dominant orbital components and symmetry quantum numbers for the highest valence band ($v$), the lowest conduction band ($c$) and the next-lowest conduction band ($c'$) at the K point in monolayer WSe$_2$ after considering the spin-preserving spin-orbit coupling term. Each band is split into two subbands with spin-up and spin-down configurations.

If we consider only the Hamiltonian with the spin-preserving $\alpha \hat{L}_z \hat{S}_z$ term, the $v$, $c$ and $c'$ bands with spin-up and spin-down configurations are the exact eigenstates. Table S3 summarizes their mirror parity and $C_3$ quantum numbers. The results are obtained by multiplying the spin and orbital mirror parity and adding the spin and orbital $C_3$ quantum numbers from Table S1 and S2.

The bright A exciton is formed by the $v_\uparrow$ and $c_\uparrow$ bands with the same spin. The dark A exciton is formed by the $v_\uparrow$ and $c_\downarrow$ bands with opposite spins. As photon cannot flip the electron spin, the dark exciton is forbidden to have radiative recombination, although such a transition is allowed by the electric-dipole selection rules.

### 3.3. Electronic bands and selection rules with spin-flipping spin-orbit coupling

For the dark exciton to emit light, it is necessary to include the spin-flipping SOC terms $\alpha(\hat{L}_+\hat{S}_- + \hat{L}_-\hat{S}_+)$ in the Hamiltonian. These two terms couple states with opposite spins and $\Delta m = \pm 1$. As the SOC obeys the crystal symmetry, only bands with the same symmetry character can be mixed together. For the $c_\downarrow$ band in the dark exciton, only the $c'_\uparrow$ band has the right spin/orbit configuration and the same symmetry character to be mixed. Other higher conduction bands (not listed in Table S1, S3) don't have the right spin/orbit and symmetry properties to couple to the $c_\downarrow$ band [8]. Also, the spin-flipping coupling of the $v_\uparrow$ band to other valence bands is very small and can be ignored here. Therefore, it is sufficient to consider only the $c'_\uparrow$ band in our analysis. Fig. S7 (c) illustrate the influence of the spin-flipping SOC on the electronic bands. As our analysis here only considers four bands ($c'_\uparrow$, $c_\uparrow$, $c_\downarrow$, $v_\uparrow$), we remove the other bands in Fig. S7(c) to simplify the presentation.

According to the first-order perturbation theory, the spin-flipping SOC will modify the $c_\downarrow$ band to be $\bar{c}_\downarrow$:

$$|\bar{c}_\downarrow\rangle = |c_\downarrow\rangle + \frac{\langle c'_\uparrow|\alpha \hat{L}_-\hat{S}_+|c_\downarrow\rangle}{E_{c_\downarrow}-E_{c'_\uparrow}}|c'_\uparrow\rangle. \quad (S5)$$

Here the "↓" subscript in $\bar{c}_\downarrow$ only means that the dominant spin component is spin-down, because it has a small spin-up component from the $c'_\uparrow$ band. As a result, the radiative recombination of dark exciton is not entirely spin-forbidden. The small $c'_\uparrow$ component will give some oscillator strength to the dark exciton so that it can absorb and emit light.

From the symmetry characters of the $\bar{c}_\downarrow$ and $v_\uparrow$ bands, we can derive the transformation of the matrix elements:

$$\langle\bar{c}_\downarrow|\hat{p}_\pm|v_\uparrow\rangle = \langle\bar{c}_\downarrow|\hat{C}_3^{-1}\hat{C}_3\hat{p}_\pm\hat{C}_3^{-1}\hat{C}_3|v_\uparrow\rangle = e^{i\frac{2\pi}{3}(C_3(\bar{c}_\downarrow)-C_3(v_\uparrow)\mp 1)}\langle\bar{c}_\downarrow|\hat{p}_\pm|v_\uparrow\rangle,$$

$$\langle\bar{c}_\downarrow|\hat{p}_\pm|v_\uparrow\rangle = \langle\bar{c}_\downarrow|\hat{\sigma}_h^{-1}\hat{\sigma}_h\hat{p}_\pm\hat{\sigma}_h^{-1}\hat{\sigma}_h|v_\uparrow\rangle = \sigma_h^*(\bar{c}_\downarrow)\sigma_h(v_\uparrow)\langle\bar{c}_\downarrow|\hat{p}_\pm|v_\uparrow\rangle,$$

$$\langle\bar{c}_\downarrow|\hat{p}_z|v_\uparrow\rangle = \langle\bar{c}_\downarrow|\hat{C}_3^{-1}\hat{C}_3\hat{p}_z\hat{C}_3^{-1}\hat{C}_3|v_\uparrow\rangle = e^{i\frac{2\pi}{3}(C_3(\bar{c}_\downarrow)-C_3(v_\uparrow))}\langle\bar{c}_\downarrow|\hat{p}_z|v_\uparrow\rangle,$$

$$\langle\bar{c}_\downarrow|\hat{p}_z|v_\uparrow\rangle = \langle\bar{c}_\downarrow|\hat{\sigma}_h^{-1}\hat{\sigma}_h\hat{p}_z\hat{\sigma}_h^{-1}\hat{\sigma}_h|v_\uparrow\rangle = -\sigma_h^*(\bar{c}_\downarrow)\sigma_h(v_\uparrow)\langle\bar{c}_\downarrow|\hat{p}_z|v_\uparrow\rangle. \quad (S6)$$

For a matrix element to be finite, the pre-factor after the symmetry transformation has to be one. From Table S3, $C_3(\bar{c}_\downarrow) - C_3(v_\uparrow) = 3N$ and $\sigma_h^*(\bar{c}_\downarrow)\sigma_h(v_\uparrow) = -1$. So, only $\langle \bar{c}_\downarrow | \hat{p}_z | v_\uparrow \rangle$ can be finite, whereas $\langle \bar{c}_\downarrow | \hat{p}_\pm | v_\uparrow \rangle$ is zero. The dark exciton is therefore coupled exclusively to the z-polarized light with out-of-plane dipole (for both the K and K' valleys).

### 3.4. Electronic bands and selection rules with electron-phonon coupling

Now let's consider the dark-exciton phonon replica. The replica effect is induced by the electron-phonon coupling Hamiltonian $\hat{H}_{ep}$. As the lattice displacement breaks the symmetry of the crystal, the original bands are no long the exact eigenstates of the new Hamiltonian with electron-phonon coupling. According to the first-order perturbation theory, $\hat{H}_{ep}$ will renormalize the $\bar{c}_\downarrow$ state to be $\bar{\bar{c}}_\downarrow$:

$$| \bar{\bar{c}}_\downarrow \rangle = | \bar{c}_\downarrow \rangle + \frac{\langle c_\uparrow, \Omega | \hat{H}_{ep} | \bar{c}_\downarrow \rangle}{E_{\bar{c}_\downarrow} - E_{c_\uparrow} - \hbar\Omega} | c_\uparrow \rangle. \tag{S7}$$

Here $| c_\uparrow, \Omega \rangle$ represents the $c_\uparrow$ state plus one E" phonon with energy $\hbar\Omega$. We only consider the $c_\uparrow$ band, which makes dominant contribution because it is very close to the $\bar{c}_\downarrow$ band. The dark exciton is mixed with the bright exciton by the electron-phonon interaction. Other bands will also go through some tiny renormalizations, which can be ignored in our analysis.

There are two crucial points concerning the electron-phonon coupling between the bright and dark excitons. First, the phonon mode needs to have odd mirror parity in order to couple the $c_\uparrow$ and $\bar{c}_\downarrow$ bands with opposite mirror parity. The E" optical phonons in monolayer WSe$_2$ fulfill this requirement because the top and bottom Se atoms move in opposite directions in the E" vibration. But for other optical phonons, in which the top and bottom Se atoms move in the same direction, they have even mirror parity and cannot couple the bright and dark excitons (so we don't see their replica emission).

Second, the coupling with the E" phonon obeys a chiral selection rule. The LO and TO branches of the E" phonons are degenerate at the zone center. In these two modes, the W atoms are stationary and the Se atoms move along either the armchair or zigzag directions. Their linear superposition can form two chiral phonon modes with right and left handedness [18]. The right-handed chiral phonon (denoted as $\Omega^+$) involves counter-clockwise rotation of the Se atoms; it has angular momentum $l = +1$. The left-handed chiral phonon (denoted as $\Omega^-$) involves clockwise rotation of the Se atoms; it has angular momentum $l = -1$. Both chiral phonons have odd mirror parity because the top and bottom Se atoms are always in different lateral positions during the rotation. They can therefore couple the bright and dark states with opposite mirror parity. Notably, the zone-center chiral phonons exhibit a three-fold rotational symmetry around the W atoms. As the $\Omega^\pm$ states have the symmetry character of $x \pm iy$, they transform as:

$$\hat{C}_3 | \Omega^\pm \rangle = e^{\mp i\frac{2\pi}{3}} | \Omega^\pm \rangle. \tag{S8}$$

The $\hat{C}_3$ quantum number for $\Omega^\pm$ is $\pm 1$ (Table S4).

| Chiral phonons | $\sigma_h$ | $C_3(W)$ |
|---|---|---|
| Right-handed phonon $\Omega^+$ ($l = +1$) | $-1$ | $+1$ |
| Left-handed phonon $\Omega^-$ ($l = -1$) | $-1$ | $-1$ |

**Table S4.** The symmetry quantum numbers for the zone-center chiral phonons in monolayer WSe$_2$.

The $C_3$ rotation symmetry will impose a selection rule in the electron-phonon coupling matrix element:

$$\langle c_\uparrow, \Omega^\pm | \hat{H}_{ep} | \bar{c}_\downarrow \rangle = \langle c_\uparrow, \Omega^\pm | C_3^{-1} C_3 \hat{H}_{ep} C_3^{-1} C_3 | \bar{c}_\downarrow \rangle$$
$$= e^{i\frac{2\pi}{3}(C_3(c_\uparrow) - C_3(\bar{c}_\downarrow) + C_3(\Omega^\pm))} \langle c_\uparrow, \Omega | \hat{H}_{ep} | \bar{c}_\downarrow \rangle$$
$$= e^{i\frac{2\pi}{3}(1 + C_3(\Omega^\pm))} \langle c_\uparrow, \Omega | \hat{H}_{ep} | \bar{c}_\downarrow \rangle. \qquad (S9)$$

Here $C_3 \hat{H}_{ep} C_3^{-1} = \hat{H}_{ep}$ because the electron-phonon coupling Hamiltonian has the three-fold rotation symmetry. From Table S3, $C_3(c_\uparrow) - C_3(\bar{c}_\downarrow) = 1$. The matrix element can be finite only when $1 + C_3(\Omega^\pm) = 3N$. This can be fulfilled only by the left-handed chiral phonon with $C_3(\Omega^-) = -1$. Therefore, the K-valley dark exciton only emits the left-handed chiral phonon.

Next, we consider the phonon-assisted radiative recombination of dark exciton through the $\bar{\bar{c}}_\downarrow - v_\uparrow$ transition with the electron-light interaction $\hat{H}_{el}$. The Fermi's golden rule gives:

$$P_{\bar{\bar{c}}_\downarrow - v_\uparrow} = \frac{2\pi}{\hbar} \left| \langle v_\uparrow, \omega, \Omega | \hat{H}_{el} | \bar{\bar{c}}_\downarrow \rangle \right|^2 \delta\left(E_{v_\uparrow} - E_{\bar{\bar{c}}_\downarrow} - \hbar\Omega - \hbar\omega\right)$$
$$= \frac{2\pi}{\hbar} \left| \frac{\langle v_\uparrow, \omega, \Omega | \hat{H}_{el} | c_\uparrow, \Omega \rangle \langle c_\uparrow, \Omega | \hat{H}_{ep} | \bar{c}_\downarrow \rangle}{E_{\bar{c}_\downarrow} - E_{c_\uparrow} - \hbar\Omega} \right|^2 \delta\left(E_{v_\uparrow} - E_{\bar{\bar{c}}_\downarrow} - \hbar\Omega - \hbar\omega\right). \qquad (S10)$$

The energy of $c_\downarrow$, $\bar{c}_\downarrow$ and $\bar{\bar{c}}_\downarrow$ are essentially the same because the spin-flipping SOC and the electron-phonon coupling only modify the energy very little. The electron-photon matrix element $\langle v_\uparrow, \omega, \Omega | \hat{H}_{el} | c_\uparrow, \Omega \rangle$ corresponds to the matrix element $\langle v_\uparrow | \hat{p} | c_\uparrow \rangle$, which we have considered before for the bright exciton transition. Therefore, the chiral-phonon replica follows the intensity as well as the selection rules of the bright exciton. It is coupled only to the right-handed light with in-plane dipole, not the z-polarized light with out-of-plane dipole. In the phonon-assisted radiative recombination, the dark exciton emits left-handed chiral phonon and right-handed photon in the K-valley. By the time-reversal symmetry, the dark exciton can emit right-handed chiral phonon and left-handed photon in the K' valley.

We have also considered the possible phonon replica of bright exciton. The corresponding transition rate is:

$$P_{c_\uparrow - v_\uparrow} = \frac{2\pi}{\hbar} \left| \frac{\langle v_\uparrow, \omega, \Omega | \hat{H}_{el} | c_\downarrow, \Omega \rangle \langle c_\downarrow, \Omega | \hat{H}_{ep} | c_\uparrow \rangle}{E_{c_\uparrow} - E_{c_\downarrow} - \hbar\Omega} \right|^2 \delta(E_{c_\uparrow} - E_{v_\uparrow} - \hbar\Omega - \hbar\omega) = 0. \quad (S11)$$

Here the matrix element $\langle v_\uparrow, \omega, \Omega | \hat{H}_{el} | c_\downarrow, \Omega \rangle = 0$ due to the spin mismatch because electron-photon coupling cannot flip the spin. Therefore, bright exciton has no chiral-phonon replica, consistent with our experimental results.

In summary, we have derived the selection rules for the bright exciton, dark exciton and phonon replica according to group theory. The bright exciton at the K(K') valley is coupled to the right-handed (left-handed) light with in-plane dipole. The dark exciton at both valleys is coupled to z-polarized light with out-of-plane dipole. The bright exciton has no chiral-phonon replica. The dark exciton has chiral-phonon replica, which follows the same selection rules of the bright exciton. That is, the dark-exciton phonon replica at the K(K') valley is coupled to the left-handed (right-handed) chiral phonon and the right-handed (left-handed) photon. Our experimental results are fully consistent with these predicted selection rules.

### 3.5. Possible photon-phonon entanglement from the decay of dark exciton

The phonon-assisted radiative recombination of dark exciton can possibly induce an entanglement state between a single photon and a chrial phonon. Recent research shows that, in the absence of magnetic field, the two degenerate valleys can mix to produce a valley-coherent dark exciton [19]. In another word, the $D^0$ state is a coherent superposition of the K-valley state and the K'-valley state at zero magnetic field. The corehent state can be expressed generally as $|K\rangle + e^{i\theta}|K'\rangle$, where $e^{i\theta}$ is a phase factor between the two valleys.

Let's now consider the phonon-assisted recombination of the $D^0$ exciton. Its K-valley component will emit a left-handed chiral phonon ($\Omega^-$) and a right-handed photon ($\sigma^+$). Its K'-valley component will emit a right-handed chiral phonon ($\Omega^+$) and a left-handed photon ($\sigma^-$) (Fig. S8). The resultant photon-phonon total state $|\Psi_{tot}\rangle$ is hence a coherent suposition of both paths:

$$|\Psi_{tot}\rangle = \frac{1}{\sqrt{2}}\left(|\Omega^-\rangle \otimes |\sigma^+\rangle + e^{i\theta}|\Omega^+\rangle \otimes |\sigma^-\rangle\right). \quad (S12)$$

This is an entangled state of a single photon and a chrial phonon. Such a photon-phonon entanglement state has been reported in the phonon-assisted quantum-dot excition emission in monolayer WSe$_2$ [18]. In principle, it can also be realized in the dark excitons in monolayer WSe$_2$. Compared to the stationary quantum dots, the mobile dark exctions have much advantage in the application of quantum information technology. Further research is merited to explore the rich physics in this direction.

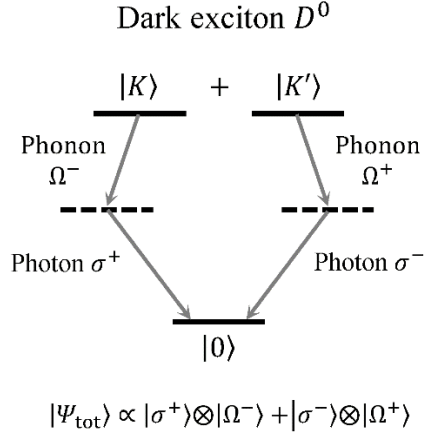

Dark exciton $D^0$

$$|\Psi_{\text{tot}}\rangle \propto |\sigma^+\rangle \otimes |\Omega^-\rangle + |\sigma^-\rangle \otimes |\Omega^+\rangle$$

**Fig. S8.** The schematic of quantum entanglement between a single photon and a chiral phonon after the phonon-assisted radiative recombination of dark exciton.

## 4. First-principles calculations of the emission intensity

While the excitonic effect does not modify the phonon and photon selection rules, it can substantially enhance the intensity of phonon-replica emission. We have calculated the relative optical strength of bright exciton, dark exciton and phonon replica by the density functional theory (DFT) using the WIEN2k package. Our first-principles calculations consider a full excitonic picture. The theoretical results are compatible with our experimental observation.

### 4.1. Intensity ratio between bright exciton and dark exciton

An exciton is formed by the linear superposition of electron-hole pairs over an envelope function. We can define the bright and dark exciton as:

$$|A\rangle = \sum_k \varphi_A(\boldsymbol{k}) | v^h_{\uparrow,\boldsymbol{k}}; c_{\uparrow,\boldsymbol{k}} \rangle,$$
$$|D\rangle = \sum_k \varphi_D(\boldsymbol{k}) | v^h_{\uparrow,\boldsymbol{k}}; \bar{c}_{\downarrow,\boldsymbol{k}} \rangle. \qquad (S13)$$

Here $| v^h_{\uparrow,\boldsymbol{k}}; c_{\uparrow,\boldsymbol{k}} \rangle$ and $| v^h_{\uparrow,\boldsymbol{k}}; \bar{c}_{\downarrow,\boldsymbol{k}} \rangle$ denote a pair of electron and hole in the conduction and valence bands, respectively, with wave vector $\boldsymbol{k}$. $\varphi_A(\boldsymbol{k})$ and $\varphi_D(\boldsymbol{k})$ are the envelope functions for bright and dark excitons, respectively. They can be calculated by the approach as described in Ref. [20]. The $\bar{c}_\downarrow$ band is the mixture of the $c_\downarrow$ and $c'_\uparrow$ bands by SOC. It has previously been obtained as:

$$|\bar{c}_\downarrow\rangle = c + f | c'_\uparrow \rangle,$$
$$\text{where } f = \frac{\langle c'_\uparrow | \alpha \hat{L}_- \hat{S}_+ | c_\downarrow \rangle}{E_{c_\downarrow} - E_{c'_\uparrow}}. \qquad (S14)$$

The matrix element $f$ is approximately independent of $\boldsymbol{k}$ near the K point. Our DFT calculation shows that $|f| \approx 0.17$. The SOC mixing of the $c_\downarrow$ and $c'_\uparrow$ bands is therefore significant. The ratio of the oscillator strength between the bright and dark exciton is calculated to be:

$$\frac{S_A}{S_D} \approx \frac{|\langle \bar{c}_\downarrow | \hat{p}_z | v_\uparrow \rangle|^2}{|\langle \bar{c}_\downarrow | \hat{p}_+ | v_\uparrow \rangle|^2} \approx 400. \tag{S15}$$

We have previously measured the lifetime of bright and dark exciton by time-resolved PL experiment [2]. The dark exciton lifetime is ~200 ps. The bright exciton lifetime is much shorter. Due to the limited time-resolution of our setup, we can only put an upper bound of 4 ps on the bright exciton lifetime. Their photoluminescence (PL) ratio is estimated to be $\frac{I_A}{I_D} < \frac{400 \times 4}{200} \approx 8$. Our experimental ratio is $I_A/I_D \approx 0.1$ after we correct the different collection efficiency of the IP and OP emission. This suggests that the bright exciton lifetime is much shorter than 4 ps. Femtosecond time-resolved experiments are required to resolve this issue.

### 4.2. Intensity ratio between dark exciton and phonon replica

The transition rate for the dark-exciton phonon replica is:

$$P(\omega) = \frac{2\pi}{\hbar} \sum_\mathbf{q} \left| \frac{\langle 0, \omega, \Omega_\mathbf{q} | \hat{H}_{el} | A, \Omega_\mathbf{q} \rangle \langle A, \Omega_\mathbf{q} | \hat{H}_{ep} | D \rangle}{E_D - E_A - \hbar\Omega_\mathbf{q}} \right|^2 \delta(E_D - \hbar\Omega_\mathbf{q} - \hbar\omega). \tag{S16}$$

Here $|0\rangle$ refers to the vacuum state of electron-hole pair. The key feature here is that we can involve a large number of phonons with finite momentum ($\Omega_\mathbf{q}$) in the matrix element $\langle A, \Omega_\mathbf{q} | \hat{H}_{ep} | D \rangle$ because of the finite k-space of the exciton envelope functions. The k-space extent of the coupled phonons is approximately the same as the k-space extent of the bright and dark exciton envelope functions. Since a tightly bound exciton has a broad envelope function, the strong excitonic effect in monolayer WSe$_2$ can substantially enhance the intensity of the phonon replica.

The chiral selection rules of phonon replica still hold even after we include the phonons near the zone center. As the electron-phonon coupling Hamiltonian has the same symmetry of the crystal, symmetry requires that the phonon states will superpose coherently and couple to the excitons in a chiral way that obeys the selection rules that we derived before. Therefore, the excitonic effect does not affect the selection rules.

We can write the electron-phonon coupling matrix element in terms of the exciton envelope functions:

$$\langle A, \Omega_\mathbf{q} | \hat{H}_{ep} | D \rangle = f \sum_\mathbf{k} \varphi_A^*(\mathbf{k}-\mathbf{q}) \varphi_D(\mathbf{k}) \langle c_{\uparrow,\mathbf{k}-\mathbf{q}}, \Omega_\mathbf{q} | \hat{H}_{ep} | c'_{\uparrow,\mathbf{k}} \rangle. \tag{S17}$$

Here we only keep the $f|c'_\uparrow\rangle$ component in the $|\bar{c}_\downarrow\rangle$ Bloch state defined in (S14), since the electron-phonon interaction cannot flip the spin. We have calculated the matrix element $\langle c_{\uparrow \mathbf{k}-\mathbf{q}}, \Omega_\mathbf{q} | \hat{H}_{ep} | c'_{\uparrow \mathbf{k}} \rangle$ by DFT. It is approximately independent of **k** in the small range near the K point. The electron-phonon interaction is: [21]

$$\hat{H}_{ep} = \Sigma_{n,j,\sigma,\mathbf{q}} \sqrt{\frac{\hbar}{2NM_j\Omega_{\sigma q}}} \nabla_j V(\mathbf{r} - \mathbf{R}_n - \mathbf{R}_j) \cdot \boldsymbol{\epsilon}_j^\sigma (a_{\sigma\mathbf{q}}^+ + a_{\sigma\mathbf{q}}) e^{i\mathbf{q}\cdot\mathbf{R}_n}. \quad (S18)$$

Here $N$ denotes the number of unit cells in the sample; $\mathbf{R}_n$ denotes the position of every unit cell in the crystal; $M_j$ and $\mathbf{R}_j$ denotes the mass and position of an atom at site $j$ in a unit cell; $V(\mathbf{r} - \mathbf{R}_n - \mathbf{R}_j)$ denotes the crystal potential (electron-ion interaction screened by the valence electrons) of the atomic site at $\mathbf{R}_n + \mathbf{R}_j$; $a_{\sigma\mathbf{q}}^+$ ($a_{\sigma\mathbf{q}}$) creates (annihilates) a phonon of polarization mode $\sigma$ and wave-vector $\mathbf{q}$; $\boldsymbol{\epsilon}_j^\sigma$ is the corresponding polarization vector. For $E''$ phonon of interest, there are two modes (longitudinal and transverse modes) at finite $\mathbf{q}$ with a small splitting. $\nabla_j$ denotes the gradient with respect to $\mathbf{R}_j$. Thus,

$$\Sigma_{j=1,2} \nabla_j V(\mathbf{r} - \mathbf{R}_j) \cdot \boldsymbol{\epsilon}_j^\sigma = \frac{1}{\sqrt{2}} \left[ \frac{\partial}{\partial R_{1\sigma}} V(\mathbf{r} - \mathbf{R}_1) - \frac{\partial}{\partial R_{2\sigma}} V(\mathbf{r} - \mathbf{R}_2) \right]. \quad (S19)$$

Here $\mathbf{R}_1$ and $\mathbf{R}_2$ denote the positions of the two Se atoms in the $\mathbf{R}_n = 0$ unit cell. $\frac{\partial}{\partial R_{j\sigma}}$ denotes the derivative with respect to $\mathbf{R}_j$ along the direction of polarization of mode $\sigma$. The matrix elements between the basis functions centered at the atomic sites can be conveniently evaluated, since the gradient on $\mathbf{R}_j$ can be transferred to the gradient on $\mathbf{r}$. We have carried out a DFT calculation by using a full potential augmented plane wave (APW) code [22] projected into Slater-type orbitals [23, 24]. We obtain the electron-phonon coupling, which is a function of $\mathbf{q}$ (when over $\mathbf{k}$ near the K point weighted by $|\varphi_D(\mathbf{k})|^2$). We write

$$|\langle c_{\uparrow,\mathbf{k}-\mathbf{q}}, \Omega_{\mathbf{q}} | \hat{H}_{ep} | c'_{\uparrow,\mathbf{k}} \rangle| = \Xi(\mathbf{q})/\sqrt{N}. \quad (S20)$$

The value of $|\Xi(q)|$ ranges from 0.013eV to 0.05eV for $q$ between 0 and 0.1 atomic unit when the full valence electron screening is included. If we assume partial screening, which is possible for interband scattering by an optical phonon, the value can change by a factor of two. By substituting these results into (S16), we yield an integrated oscillator strength (S) for the dark-exciton phonon replica:

$$S_{Dp} = \int d\omega\, P(\omega) \propto \frac{1}{N} |\langle 0, \omega | \hat{H}_{el} | A \rangle|^2 \Sigma_{\mathbf{q}} \left| \frac{\Xi(\mathbf{q}) f}{E_D - E_A - \hbar\Omega_q} O_{AD}(\mathbf{q}) \right|^2. \quad (S21)$$

Here $O_{AD}(\mathbf{q}) = \Sigma_{\mathbf{k}} \varphi_A^*(\mathbf{k}-\mathbf{q}) \varphi_D(\mathbf{k})$ denotes the Fourier transform of the product of real-space envelope functions for the dark and bright excitons. The integral over $|O_{AD}(\mathbf{q})|^2$ is inversely proportional to the square of the exciton radius.

In comparison, the integrated oscillator strength for the dark (bright) exciton zero-phonon peak is:

$$S_A \propto |\langle 0, \omega | \hat{H}_{el} | A \rangle|^2,$$
$$S_D \propto |\langle 0, \omega | \hat{H}_{el} | D \rangle|^2. \quad (S22)$$

Finally, by using $|E_D - E_A - \hbar\Omega_q| \approx 0.062$ eV and carrying out the integral in (S21),

we obtain $S_{D_p^0}/S_{A^0} \approx 8 \times 10^{-5}$ ($2 \times 10^{-5}$) and $S_{D_p^0}/S_{D^0} \approx 0.016$ (0.004) for the condition of partial screening (full screening).

In our experiment, the measured intensity ratio between the phonon replica and dark exciton is $S_{D_p^0}/S_{D^0} \approx 0.5$, which is much larger than the calculated ratio. However, we must consider the different collection efficiency of the IP and OP emission in our measurement geometry. We assume that the radiation intensity follows the $\sin^2\theta$ pattern predicted in classical electrodynamics for an oscillating dipole. Our objective has a numerical aperture of 0.6. For the replica with IP dipole, the radiation concentrates at small angles. The collection efficiency is ~60%. For the dark exciton with OP dipole, the radiation concentrates at large angles. The collection efficiency is only ~7%. We have included the reflected PL from the silicon substrate in these estimations. After considering the different collection efficiency, the ratio of oscillator strength between the phonon replica and dark exciton is estimated to be $S_{D_p^0}/S_{D^0} \approx 0.055$. The experimental ratio is in the same order of magnitude as our predicted ratio $S_{D_p^0}/S_{D^0} \approx 0.016$ for partial screening.

## Supplemental References: